\documentclass{elsarticle}
\usepackage{graphicx,amssymb}
\journal{Physica A}

\begin{document}

\begin{frontmatter}



\title{Reentrant phase transitions and multicompensation points 
in the mixed-spin Ising ferrimagnet on a decorated Bethe lattice}

\author[UPJS]{Jozef Stre\v{c}ka\fnref{ack2}} 
\ead{jozef.strecka@upjs.sk}
\address[UPJS]{Department of Theoretical Physics and Astrophysics, Faculty of Science, \\
P. J. \v{S}af\'{a}rik University, Park Angelinum 9, 040 01 Ko\v{s}ice, Slovak Republic}
\author[AMU]{Cesur Ekiz\fnref{ack1}}
\ead{ekizc@yahoo.com}
\address[AMU]{Department of Physics, Faculty of Science, Adnan Menderes University, \\
09010 Ayd\i n, Turkey}

\begin{abstract}
Mixed-spin Ising model on a decorated Bethe lattice is rigorously solved by combining the decoration-iteration transformation 
with the method of exact recursion relations. Exact results for critical lines, compensation temperatures, total and sublattice magnetizations are obtained from a precise mapping relationship with the corresponding spin-1/2 Ising model on a simple (undecorated) Bethe lattice. The effect of next-nearest-neighbour interaction and single-ion anisotropy on magnetic properties of the ferrimagnetic model is investigated in particular. It is shown that the total magnetization may exhibit multicompensation phenomenon and the critical temperature vs. the single-ion anisotropy dependence basically changes with the coordination number of the underlying Bethe lattice. The possibility of observing reentrant phase transitions is related to a high enough coordination number of the underlying Bethe lattice.
\end{abstract}


\end{frontmatter}

\section{Introduction}

Ferrimagnetism has been extensively studied in the past both experimentally as well as theoretically. The magnetic structure of insulating ferrimagnets is built up from several (at least two) sublattices, which are occupied by different carriers of unequal magnetic moments interacting with each other through the antiferromagnetic interaction. Mixed-spin Ising models probably represent the simplest theoretical models, which may bring an insight into the most important features accompanying ferrimagnetic ordering such as for instance a compensation phenomenon. A compensation point denotes the temperature $T_{comp}$, at which the resultant magnetization vanishes due to a complete cancellation of sublattice magnetizations below the critical temperature $T_{c}$, i.e. $T_{comp} < T_{c}$ \cite{Neel}. An existence of compensation temperatures has obvious technological significance in the field of thermomagnetic recording, because only a small driving field is needed at a compensation point in order to achieve a magnetic pole reversal \cite{Mansur, Cornel}. Besides, the mixed-spin Ising models belong to the most interesting extensions of the standard spin-1/2 Ising model, which may display more diverse critical behaviour compared with their single-spin counterparts.

Decorated mixed-spin Ising models, which were originally introduced by Syozi and co-workers (see Ref. \cite{Syoz} and references cited therein), have been intensively studied many years ago as theoretical models of ferrimagnetism. Thermal variations of the magnetization in the decorated Ising ferrimagnets have been investigated in detail and these results shed light on some characteristic features of real ferrimagnetic materials \cite{Naka,Hatto}. The effect of single-ion anisotropy on magnetic properties of the mixed spin-1/2 and spin-$S$ ($S>1/2$) Ising ferrimagnet on decorated planar lattices has been examined by Kaneyoshi with the help of differential operator technique and the effective-field theory \cite{Kan96,Kan97}. By making use of these theoretical tools, Kaneyoshi has predicted many interesting features of the decorated Ising ferrimagnets such as an appearance of two compensation points and/or multiple reentrant phase transitions. It is worthwhile to remark that Kaneyoshi's predictions were subsequently confirmed by Ja\v{s}\v{c}ur \cite{Jascur98} and Dakhama \cite{Dakhama98} when solving the ferrimagnetic version of the decorated Ising model rigorously. The significance of this model also lies in the fact that temperature dependences of the total magnetization show numerous characteristic features not predicted in the standard N\'eel theory of ferrimagnetism \cite{Neel}.

Recently, the magnetic properties of various decorated mixed-spin Ising models have been explored in detail by variety of mathematical techniques with some exact \cite{Jascur98,Dakhama98,Jascur99,Strecka03,Oitmaa03,Strecka07,Matasovska07,Canova08} as well as approximate results \cite{Kaney96,Eddeqaqi99,Kan01,Kan02,Moutie02,Benyo07,Saber07,Boughrara08,Elkenz09,Dely09}. The present work deals with magnetic properties of the mixed-spin Ising ferrimagnet on the decorated Bethe lattice accounting for the next-nearest-neighbor interaction and the uniaxial single-ion anisotropy. The exact solution will be obtained by combining the decoration-iteration transformation with the another rigorous method based on recursion relations. First, the decoration-iteration transformation will be applied to establish the precise mapping correspondence between the mixed-spin Ising model on the decorated Bethe lattice and its equivalent spin-1/2 Ising model on the simple (undecorated) Bethe lattice. It is well known that this latter model can be subsequently treated by making use of recursion relations \cite{baxt82}, which will help us to complete our exact calculation for the original mixed-spin 
Ising model on the decorated Bethe lattice. Within the framework of these two exact methods, we will focus on the effect of the single-ion anisotropy and the next-nearest-neighbour interaction on the phase diagrams, critical behaviour and compensation phenomenon of the model under investigation.

The outline of the present paper is as follows. In Section 2, the detailed description of the model system will be presented and the basic steps of both exact methods will be briefly clarified. The most interesting results are presented and discussed in detail in Section 3. In particular, our attention is focused on the finite-temperature phase diagrams, reentrant phase transitions and temperature variations of both sublattice as well as total magnetizations. Finally, some conclusions are mentioned in the last Section IV.

\section{Model and method}

\begin{figure}
\begin{center}
\includegraphics[width=0.8\textwidth]{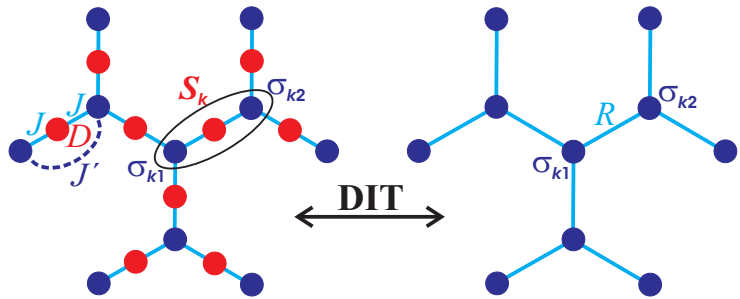}
\end{center}
\vspace{-0.2cm}
\caption{The mixed spin-1/2 and spin-$S$ Ising model on the decorated Bethe lattice (figure on the left) and its relation to the corresponding spin-1/2 Ising model on the simple (undecorated) Bethe lattice with the coordination number $q=3$ (figure on the right). The blue (red) circles denote lattice positions of the spin-1/2 (spin-1) Ising spins and the ellipse demarcates all the interaction terms belonging to the $k$th bond Hamiltonian.}
\label{fig1}
\end{figure}

Let us begin by considering the mixed spin-1/2 and spin-$S$ Ising model on the decorated Bethe lattice with a quite general coordination number $q$ as schematically illustrated in Fig. \ref{fig1} on the particular example of the Bethe lattice with the coordination number $q=3$. The magnetic structure of the investigated model system constitutes the Ising spins $\sigma=1/2$ placed on lattice sites of a deep interior of infinite Cayley tree (Bethe lattice) and the Ising spins of a quite general magnitude $S$ placed on each bond of the original Bethe lattice. The total Hamiltonian of the mixed-spin Ising model on the decorated Bethe lattice then reads
\begin{equation}
{\cal H} = -J\sum_{i, j}^{nn} S_{i}\sigma_{j} -{J}'\sum_{k,j}^{nnn} \sigma_{k} \sigma_{j}-D\sum_{i=1}^{Nq/2}S_{i}^{2},
\label{eq1}
\end{equation}
where $\sigma_{j}=\mp1/2$ and $S_{i} = -S,-S+1,\ldots,S$, the relevant subscript specifies the lattice position and the exchange constants $J$ and $J'$ represent the nearest-neighbour interaction between the spin-1/2 and spin-$S$ atoms and, respectively, the next-nearest-neighbour interaction between the spin-1/2 atoms. Further, the parameter $D$ stands for the uniaxial single-ion anisotropy acting on the spin-$S$ atoms only. As one can see from Fig.~\ref{fig1}, the magnetic structure of the investigated model system constitutes two interpenetrating sublattices. The former sublattice A is formed by the sites of original (undecorated) Bethe lattice 
that is occupied by the atoms with the fixed spin $\sigma=1/2$, while the latter sublattice B is occupied by the decorating atoms 
with an arbitrary spin value $S$. The calculation on the Bethe lattice is done recursively \cite{baxt82}.

Now, we will turn our attention to main points of the transformation method, which provides a rigorous mapping equivalence between the investigated spin model and its corresponding spin-1/2 Ising model on undecorated Bethe lattice. First, it is quite useful to rewrite the total Hamiltonian (\ref{eq1}) as a sum of the bond Hamiltonians, i.e. ${\cal H}=\sum_{k=1}^{Nq/2} {\cal H}_{k}$, where $N$ denotes the total number of the spin-1/2 atoms from the sublattice A, $q$ is the coordination number of the undecorated Bethe lattice and the summation runs over all bonds of the decorated Bethe lattice. Thus, each bond Hamiltonian ${\cal H}_{k}$ involves all the interaction terms including the decorating Ising spin $S_k$ and the next-nearest-neighbour interaction between its two nearest-neighbouring Ising spins $\sigma_{k1}$ and $\sigma_{k2}$ (for better illustration see spin cluster enclosed by ellipse in Fig. \ref{fig1})
\begin{equation}
{\cal H}_k =-JS_{k}(\sigma _{k1}+\sigma _{k2})-DS_{k}^{2}-J'\sigma _{k1}\sigma _{k2}. 
\label{eq2}
\end{equation}
A crucial step in our procedure represents the evaluation of the partition function. For further convenience, the partition function of the model under consideration can be written in the following useful form
\begin{eqnarray}
{\cal Z} = \sum_{\{\sigma_{i}\}}\prod_{k = 1}^{Nq/2} \sum_{S_{k}}{\exp(-\beta {\cal H}_k)},
\label{eq3}
\end{eqnarray}
where $\beta=1/(k_{B}T)$, $k_{B}$ is the Boltzmann's constant and $T$ is the absolute temperature. The symbol $\sum_{S_{k}}$ 
denotes a summation over $2S+1$ spin states of the $k$th decorating Ising spin from the sublattice B, while the symbol $\sum_{\{\sigma_{i}\}}$ marks a summation over all available spin configurations of the complete set of the spin-1/2 Ising 
atoms from the sublattice A. 
It is noteworthy that the summation over spin states of the decorating spin-$S$ atoms can be performed independently of each other
because of the lack of any direct interaction among them. After performing the summation over the spin states of the decorating Ising spin $S_k$, one gets the effective Boltzmann's weight that implies a possibility of applying 
the generalized decoration-iteration transformation \cite{Fisher,onofre,jozef}
\begin{eqnarray}
\sum_{S_{k}} \exp(-\beta {\cal H}_k) \!\!\!&=&\!\!\! \exp( \beta J' \sigma_{k1} \sigma_{k2}) 
\sum_{n=-S}^{S} \exp(\beta D n^2) \cosh[\beta J n (\sigma_{k1} + \sigma_{k2}) ] \nonumber \\
\!\!\!&=&\!\!\! A \exp( \beta R \sigma_{k1} \sigma_{k2}).
\label{eq4}
\end{eqnarray}
The physical meaning of the mapping transformation (\ref{eq4}) lies in removing all the interaction parameters associated with the  decorating spin $S_k$ and replacing them by a new unique effective interaction $R$ between its two nearest-neighbouring Ising spins $\sigma_{k1}$ and $\sigma_{k2}$. It is worthwhile to mention that the both mapping parameters $A$ and $R$ are 'self-consistently' given directly by the transformation formula (\ref{eq4}), which must hold for all four possible spin combinations of two Ising spins $\sigma_{k1}$ and $\sigma_{k2}$ that provide just two independent equations from the mapping transformation (\ref{eq4}). In the consequence of that, two yet unknown mapping parameters $A$ and $R$ can be unambiguously determined by the following formulas
\begin{eqnarray}
A = (V_{1}V_{2})^{\frac{1}{2}}, \qquad \beta R = \beta J'+2\ln\left(\frac{V_{1}}{V_{2}}\right),
\label{eq5}
\end{eqnarray}
which contain two newly defined functions
\begin{eqnarray}
V_{1} = \sum_{n=-S}^{S} \exp(\beta D n^2) \cosh(\beta J n), \qquad
V_{2} = \sum_{n=-S}^{S} \exp(\beta D n^2).
\label{eq6}
\end{eqnarray}

After straightforward substitution of the mapping transformation (\ref{eq4}) into the formula (\ref{eq3}), one easily obtains an exact relation between the partition function of the mixed-spin Ising model on the decorated Bethe lattice and respectively, the partition function of the spin-1/2 Ising model on the simple (undecorated) Bethe lattice with the temperature-dependent effective interaction $R$
\begin{equation}
{\cal Z} (\beta, J, J', D, S, q) = A^{Nq/2} {\cal Z}_{\rm Bethe} (\beta R). 
\label{eq7}
\end{equation}
The mapping relation (\ref{eq7}) between both partition functions represents a central result of our calculation, because this relationship can in turn be employed for a rigorous calculation of some important quantities such as sublattice magnetizations and free energy. It is worthwhile to recall that this mapping correspondence is valid for arbitrary values of the decorating spin $S$ and also independently of the lattice coordination number $q$. 

Exact mapping theorems developed by Barry \textit{et al}. \cite{barry1,barry2,barry3,barry4} in combination with Eq. (\ref{eq7}) yield a straightforward relation between the spontaneous magnetization of the sublattice A of the mixed-spin Ising model on the decorated Bethe lattice and respectively, the spontaneous magnetization of the corresponding spin-1/2 Ising model on the undecorated Bethe lattice   
\begin{eqnarray}
m_{A} \equiv \frac{1}{2} \left( \langle \sigma_{k1} \rangle + \langle \sigma_{k2} \rangle \right) 
 = \frac{1}{2} \left( \langle \sigma_{k1} \rangle + \langle \sigma_{k2} \rangle \right)_{\rm Bethe}
\equiv m_{\rm Bethe} (\beta R).  
\label{eq8a}
\end{eqnarray} 
Here, the symbols $\langle \cdots \rangle $ and $\langle \cdots \rangle_{{\rm Bethe}}$ mean the standard canonical ensemble average performed in the mixed-spin Ising model on the decorated Bethe lattice defined through the Hamiltonian (\ref{eq1}) and respectively, its corresponding spin-1/2 Ising model on the undecorated Bethe lattice. According to Eq. (\ref{eq8a}), the sublattice magnetization $m_{A}$ directly equals to the spontaneous magnetization $m_{\rm Bethe}$ of the corresponding spin-1/2 Ising model on the simple Bethe lattice provided that Eqs. (\ref{eq5})-(\ref{eq6}) lead to the effective ferromagnetic interaction $R>0$. In the case of the effective antiferromagnetic interaction $R<0$, one alternatively gains a quite similar exact result for the staggered magnetization $m_A^s$ of the sublattice A of the mixed-spin Ising model on the decorated Bethe lattice
\begin{eqnarray}
m_{A}^s \equiv \frac{1}{2} \left( \langle \sigma_{k1} \rangle - \langle \sigma_{k2} \rangle \right) 
 = \frac{1}{2} \left( \langle \sigma_{k1} \rangle - \langle \sigma_{k2} \rangle \right)_{\rm Bethe}
\equiv m_{\rm Bethe} (\beta |R|), 
\label{eq8b}
\end{eqnarray}
which can also be expressed only in terms of the spontaneous magnetization $m_{\rm Bethe}$ of the corresponding spin-1/2 Ising model on the undecorated Bethe lattice. The other spontaneous sublattice magnetization of the mixed-spin Ising model on the decorated Bethe lattice can be descended from the generalized Callen-Suzuki identity \cite{call,suzu,balci}, which enables to relate it to the previously derived spontaneous sublattice magnetization $m_{A}$
\begin{eqnarray}
m_{B} \equiv \left \langle S_{k} \right \rangle = 2 m_{A} \,
\frac{\displaystyle \sum_{n=-S}^{S} n \exp(\beta D n^{2}) \sinh(\beta J n)}
     {\displaystyle \sum_{n=-S}^{S} \exp(\beta D n^{2}) \cosh(\beta J n)}.
\label{eq9}
\end{eqnarray}  

To complete our calculation of spontaneous sublattice magnetizations, one still needs the exact closed-form formula for
the spontaneous magnetization $m_{\rm Bethe}$ of the corresponding spin-1/2 Ising model on the simple (undecorated) 
Bethe lattice. This quantity can be rigorously found within the framework of exact recursion relations \cite{baxt82}. 
If the undecorated Bethe lattice (see Fig. 1) is 'cut' at a central site with the spin $\sigma_{k1}$, it will consequently 
split into $q$ identical branches. The partition function of the spin-1/2 Ising model on the simple Bethe lattice will take the form 
\begin{eqnarray}
{\cal Z}_{\rm Bethe} = \sum_{\sigma_{k1}}\left [ g_{n}(\sigma_{k1} ) \right ]^{q}=\left [ g_{n} \left(\frac{1}{2} \right) \right ]^{q}
+\left [ g_{n} \left(-\frac{1}{2}\right) \right ]^{q},
\label{eq10}
\end{eqnarray} 
where $g_{n}(\sigma_{k1})$ is the partition function of the one individual branch and $n$ is the total number of generations in the recursively
built Bethe lattice. Cutting each branch at the subsequent site $\sigma_{k2}$ allows one to find the recursive relation between the partition function 
of the branch $g_{n}(\sigma_{k1})$ with $n$ generations of spins and respectively, the partition function of the sub-branch $g_{n-1}(\sigma_{k2})$ with $n-1$ generations of spins 
\begin{eqnarray}
g_{n}(\sigma _{k1})=\sum_{\sigma_{k2} } \exp (\beta R \sigma _{k1} \sigma _{k2})\left [ g_{n-1}(\sigma_{k2}) \right ]^{q-1}. 
\label{eq11}
\end{eqnarray} 
The canonical ensemble average of the central spin $\sigma_{k1}$ in the spin-1/2 Ising model on the Bethe lattice can readily be calculated from the relation
\begin{eqnarray}
m_{\rm Bethe} = \langle \sigma_{k1} \rangle_{\rm Bethe} = \frac{1}{{\cal Z}_{\rm Bethe}} \sum_{\sigma_{k1}} \sigma_{k1} \left [ g_{n}(\sigma_{k1} ) \right ]^{q} 
= \frac{1}{2} \left( \frac{1 - x_n^q}{1 + x_n^q} \right),
\label{eq12}
\end{eqnarray} 
which contains the newly defined parameter $x_{n}=\frac{g_{n}(-\frac{1}{2})}{g_{n}(\frac{1}{2})}$. Even though the parameter $x_{n}$ does not have a 
direct physical meaning, it directly determines the overall thermodynamics of the spin-1/2 Ising model on the Bethe lattice in the limit $n \to \infty$. 
With the help of Eq. (\ref{eq11}), one may easily derive for $x_n$ that enters the magnetization formula (\ref{eq12}) the following recursion relation      
\begin{eqnarray}
x_n = \frac{\exp \left(- \frac{\beta R}{4}\right) + \exp \left(\frac{\beta R}{4}\right) x_{n-1}^{q-1}}
           {\exp \left(\frac{\beta R}{4}\right) + \exp \left(- \frac{\beta R}{4}\right) x_{n-1}^{q-1}},
\label{eq13}
\end{eqnarray} 
which can be regarded as the equation of state. The equation (\ref{eq13}) forms an iteration sequence and converges to stable fixed points in the thermodynamic limit for non-staggered phases. In our case this situation appears if the effective nearest-neighbour coupling of the corresponding spin-1/2 Ising model on the Bethe lattice is ferromagnetic (i.e. $R>0$). On the other hand, the recursion relation (\ref{eq13}) leads to two-cycle double points in the case of staggered antiferromagnetic phase, which emerges in the corresponding spin-1/2 Ising model on the Bethe lattice provided that the effective interaction is antiferromagnetic (i.e. $R<0$) \cite{vannimenus,akheyan,arakelyan,ohanyan}. However, one may take advantage of the bipartite nature of the Bethe lattice to overcome this difficulty under the simultaneous change of the nearest-neighbour interaction $R \to -R$ and the spins from each second generation $\sigma_i \to (-1)^{i+1} \sigma_i$. Hence, it follows that the staggered magnetization (\ref{eq8b}) may alternatively be calculated from the relevant exact result (\ref{eq12}) for the ferromagnetic spin-1/2 Ising model on the Bethe lattice with the effective interaction $|R|$. In this way, our exact calculation of all spontaneous sublattice magnetizations of the mixed-spin Ising model on the decorated Bethe lattice is completed, because it is now sufficient to substitute Eqs. (\ref{eq12})-(\ref{eq13}) into the formerly derived expressions (\ref{eq8a})-(\ref{eq9}) for the sublattice magnetizations $m_A$, $m_A^{s}$ and $m_B$. In doing so, one should bear in mind that the effective temperature-dependent coupling $\beta R$ given by Eqs. (\ref{eq5})-(\ref{eq6}) must enter into the recurrence relation (\ref{eq13}) in order to gain the correct exact result for the spontaneous magnetization of the corresponding spin-1/2 Ising model on the simple Bethe lattice given by Eq. (\ref{eq12}). 

Finally, let us examine in detail the critical and compensation behaviour of the model under investigation. It can be readily understood from the mapping relation (\ref{eq7}) between both partition functions that the mixed-spin Ising model on the decorated Bethe lattice may exhibit a critical point only if the corresponding spin-1/2 Ising model on the undecorated Bethe lattice is at a critical point as well. This result is taken to mean that the critical temperature of the mixed-spin Ising model on the decorated Bethe lattice can easily be obtained by comparing the effective nearest-neighbour coupling of the corresponding spin-1/2 Ising model on the simple (undecorated) Bethe lattice with its critical value $\beta_{c} |R| = 2 \ln (\frac{q}{q-2})$. The critical condition for the the mixed-spin Ising model on the decorated Bethe lattice then reads
\begin{eqnarray}
\exp \left(\frac{\beta_c J'}{2} \right) \frac{\displaystyle \sum_{n=-S}^{S} \exp(\beta_c D n^{2}) \cosh(\beta_c J n)}
     {\displaystyle \sum_{n=-S}^{S} \exp(\beta_c D n^{2})} = \left( \frac{q}{q-2} \right)^{\pm 1},
\label{eqcc}
\end{eqnarray}
where $\beta_c = 1/(k_{\rm B} T_c)$, $T_c$ denotes the critical temperature and the plus (minus) sign applies when the effective coupling $\beta R$ given by Eqs. (\ref{eq5})-(\ref{eq6}) is being positive (negative). The sign ambiguity to emerge in the critical condition (\ref{eqcc}) comes from the fact that the critical temperature of the spin-1/2 Ising model on the undecorated Bethe lattice remains the same independently of whether the effective interaction is being ferromagnetic ($R>0$) or antiferromagnetic ($R<0$). It is worthy to notice, moreover, that the nearest-neighbour spin-1/2 Ising atoms from the sublattice A should consequently exhibit ferromagnetic (antiferromagnetic) spin alignment in the phases delimited by critical phase boundaries (\ref{eqcc}) with plus (minus) sign.

Last but not least, the compensation phenomenon may also come into play if the total magnetization of the mixed-spin Ising model disappears at temperature(s) below the critical point. To explore this effect, let us therefore define the total magnetization per one spin $m_T = (2 m_A + q m_B)/(1 + q S)$, which is normalized with respect to its saturation value $m_s = (1 + q S)/(2 + q)$. In this respect, one easily finds the necessary (but not sufficient) condition for an appearance of the compensation point(s)  
\begin{eqnarray}
1 + q \frac{\displaystyle \sum_{n=-S}^{S} n \exp(\beta_{comp} D n^{2}) \sinh(\beta_{comp} J n)}
           {\displaystyle \sum_{n=-S}^{S} \exp(\beta_{comp} D n^{2}) \cosh(\beta_{comp} J n)} = 0,
\label{eqco}
\end{eqnarray}   
whereas $\beta_{comp} = 1/(k_{\rm B} T_{comp})$ and $T_{comp}$ labels the compensation temperature. It is quite obvious from Eq. (\ref{eqco}) that the compensation point may possibly arise only in the ferrimagnetic model with the antiferromagnetic ($J<0$) nearest-neighbour interaction between the spin-1/2 and spin-$S$ atoms. In addition, it also turns out from Eq. (\ref{eqco}) that the compensation temperature does not apparently depend on a strength of the next-nearest-neighbour interaction $J'$ between the spin-1/2 Ising atoms, which merely determines a magnitude of the critical temperature according to Eq. (\ref{eqcc}).  
 
\section{Results and Discussion}

Let us proceed to a discussion of the most interesting results. Even though all the results derived in the foregoing section are valid for arbitrary value of the coordination number $q$ of the underlying Bethe lattice, regardless of whether the interactions $J$ and $J'$ are ferromagnetic or antiferromagnetic and also irrespective of the value of the decorating spins, our further analysis will be henceforth restricted only to the particular case of the ferrimagnetic mixed spin-1/2 and spin-1 Ising model on the decorated Bethe lattice with a quite general coordination number $q$, the antiferromagnetic $(J<0)$ nearest-neighbour interaction and either ferromagnetic $(J'>0)$ or antiferromagnetic $(J'<0)$ next-nearest-neighbour interaction. It is noteworthy that the investigated model displays all general features of ferrimagnetism found in the mixed spin-(1/2,$S$) Ising models with the integer-valued decorating spins $S$, while the respective behaviour of the mixed spin-(1/2,$S$) Ising models with the half-odd-integer decorating spins $S$ is much less pronounced as convincingly evidenced in Refs. \cite{Kan96,Kan97,Jascur98,Dakhama98} and will be therefore postponed from the present study. Note furthermore that exact results presented hereafter may also be confronted with several known rigorous results, which have been previously obtained for the mixed spin-1/2 and spin-1 Ising ferrimagnets on the decorated planar lattices \cite{Jascur98,Dakhama98}, 
or Bethe lattices \cite{albayrak03,ekiz05,ekiz06}. To reduce the number of free parameters, all the interaction constants will be hereafter normalized with respect to a strength of the nearest-neighbour interaction $|J|$, which means that the ratio $J'/|J|$, $D/|J|$ and $k_{\rm B} T/|J|$ will accordingly measure a relative magnitude of the next-nearest-neighbour interaction, single-ion anisotropy and temperature, respectively.

\begin{figure}
\begin{center}
\includegraphics[width=0.65\textwidth]{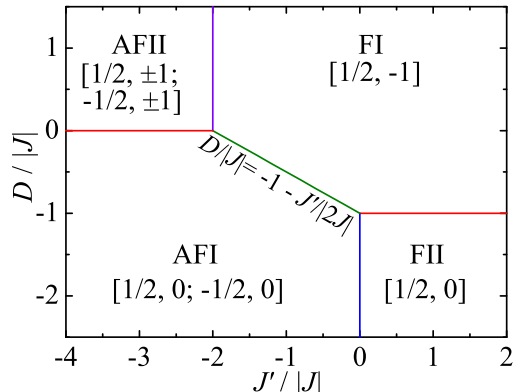}
\end{center}
\vspace{-1.2cm}
\caption{Ground-state phase diagram of the mixed spin-1/2 and spin-$1$ Ising model on the decorated Bethe lattice 
in the $J'/|J|-D/|J|$ plane. Square brackets $[\sigma, S]$ demonstrate the relevant spin alignment in each sector 
of the ground-state phase diagram. For the more detailed description of individual phases see the text.}
\label{fig2}
\end{figure}

Let us begin with the analysis of the ground state. The ground-state phase diagram of the mixed spin-1/2 and spin-$1$ Ising ferrimagnet
on the decorated Bethe lattice of arbitrary coordination number $q$ is displayed in Fig. \ref{fig2}. In agreement with our expectations, the typical ferrimagnetic phase (FI) with the antiparallel alignment $[\sigma_i, S_i] = [1/2, -1]$ between the nearest-neighbour spin-1/2 and spin-1 atoms is being the ground state provided that a relative strength of the easy-plane (negative) single-ion anisotropy $D/|J|$ and/or the antiferromagnetic next-nearest-neighbour interaction $J'/|J|$ is not too strong. 
On assumption that the next-nearest-neighbour interaction is being ferromagnetic ($J'/|J|>0$), the decrease in the single-ion anisotropy causes a first-order phase transition from FI towards the remarkable ferromagnetic phase (FII) with the respective spin alignment 
$[\sigma_i, S_i] = [1/2, 0]$. Note that the spontaneous long-range order inherent to FII is merely kept by the ferromagnetic next-nearest-neighbour interaction between the spin-1/2 atoms, whereas all the spin-1 atoms are non-magnetic as a result of sufficiently strong easy-plane single-ion anisotropy ($D/|J|<-1.0$) that energetically favours the non-magnetic spin state $S_i = 0$ before two magnetic ones $S_i = \pm 1$. If one considers the antiferromagnetic next-nearest-neighbour interaction from the intermediate range $J'/|J| \in (-2.0,0.0)$, the antiferromagnetic phase (AFI) characterized through a more complex spin arrangement $[\sigma_i, S_i; \sigma_{i+1}, S_{i+1}] = [1/2, 0; -1/2, 0]$ arises from FI upon lowering the single-ion anisotropy. All the decorating spin-1 atoms reside in AFI the non-magnetic spin state $S_i = 0$ quite similarly as in FII, but the spin-1/2 atoms from the nearest-neighbour shells of the decorated Bethe lattice align antiparallel with respect to each other on account of the antiferromagnetic nature of the next-nearest-neighbour interaction being responsible for the symmetry breaking (period doubling). It should be noted here that the emergence of both FII and AFI is restricted to the parameter space with the easy-plane character of the single-ion anisotropy $D/|J|<0$. Considering the easy-axis single-ion anisotropy $D/|J|>0$ and reinforcing the antiferromagnetic next-nearest-neighbour interaction, the mixed-spin system undergoes at $J'/|J|=-2.0$ a first-order phase transition from FI towards the partially antiferromagnetically ordered and partially disordered phase (AFII) characterized through the spin arrangement $[\sigma_i, S_i; \sigma_{i+1}, S_{i+1}] = [1/2, \pm 1; -1/2, \pm 1]$. The peculiar coexistence of a partial order and a partial disorder in AFII comes from the antiferromagnetic alignment of the next-nearest-neighbouring spin-1/2 atoms, which is simultaneously responsible for a unusual spin frustration of the decorating spin-1 atoms. Hence, it follows that the spin-1 atoms reside in AFII with the same probability one of its two magnetic spin states $S_i = \pm 1$ due to the easy-axis character of the single-ion anisotropy $D/|J|>0$. 
It should be pointed out that the established ground-state phase diagram holds for the mixed-spin Ising models on the decorated planar lattices as well. It surprisingly appears that the characterization of the phase FII has been notably overlooked in several previous approximative \cite{Kan96,Kan97} as well as exact \cite{Jascur98,Dakhama98} studies on the mixed-spin Ising model on decorated planar lattices. In addition, it seems that the particular case with the antiferromagnetic next-nearest-neighbour interaction has not been dealt with hitherto and hence, theoretical predictions of these two phases represent the completely new finding not reported in the literature yet.

\begin{figure}
\includegraphics[width=0.53\textwidth]{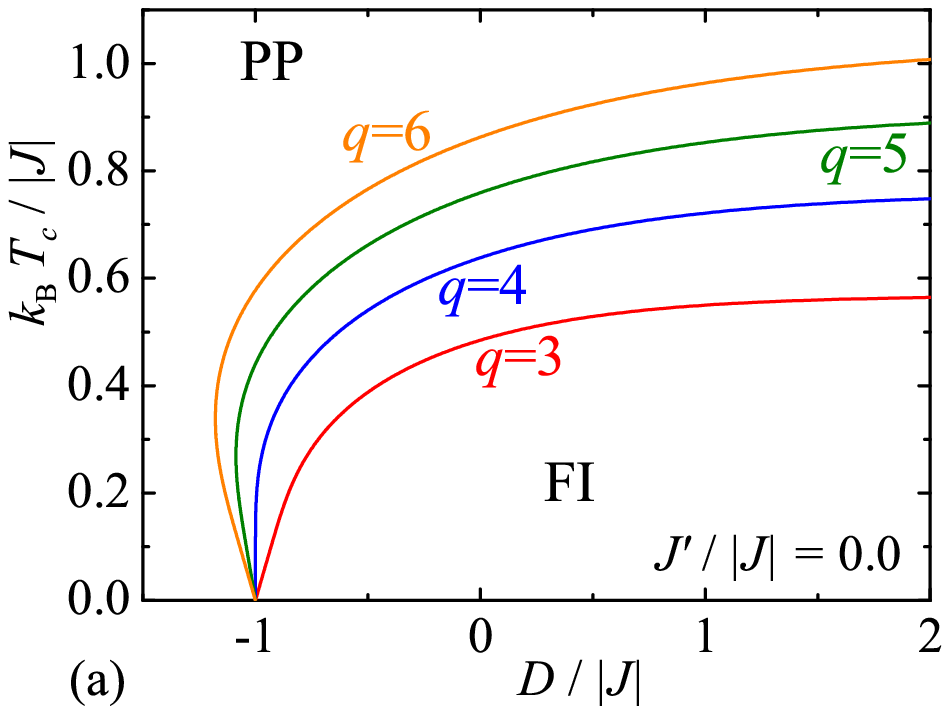}
\hspace{-0.6cm}
\includegraphics[width=0.53\textwidth]{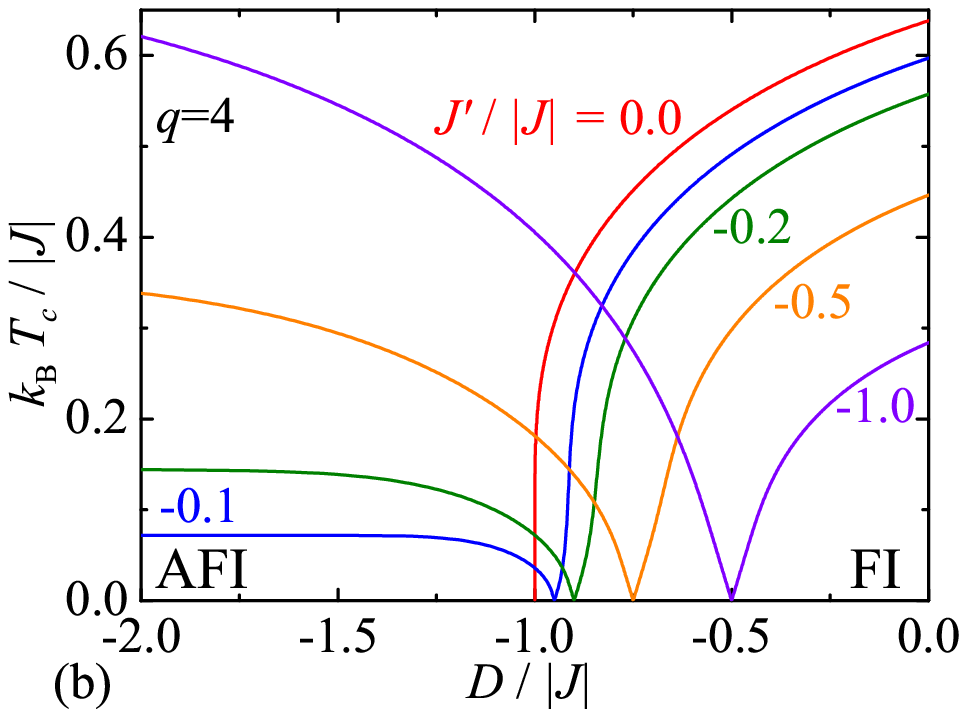}
\vspace{-1cm}
\caption{Critical temperature of the mixed spin-1/2 and spin-1 Ising model on the decorated Bethe lattice as a function of the single-ion anisotropy $D/|J|$ for: (a) $J'/|J|=0.0$ and $q$=3, 4, 5 and 6; (b) $q=4$ and $J'/|J| \leq 0$.}
\label{fig3}
\end{figure}

\begin{figure}
\includegraphics[width=0.53\textwidth]{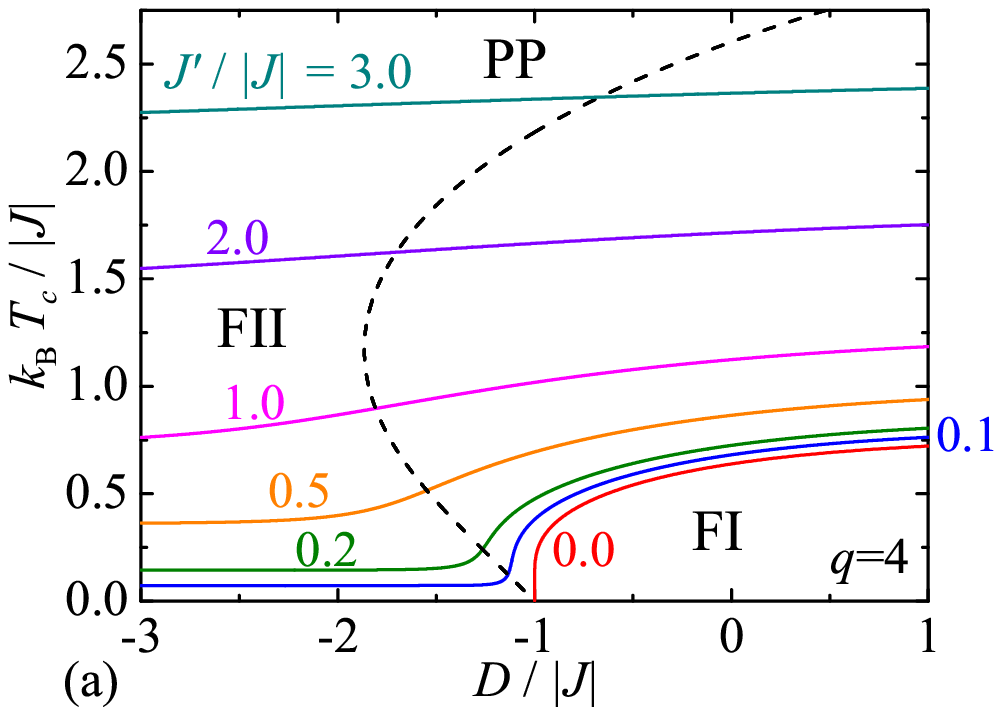}
\hspace{-0.6cm}
\includegraphics[width=0.53\textwidth]{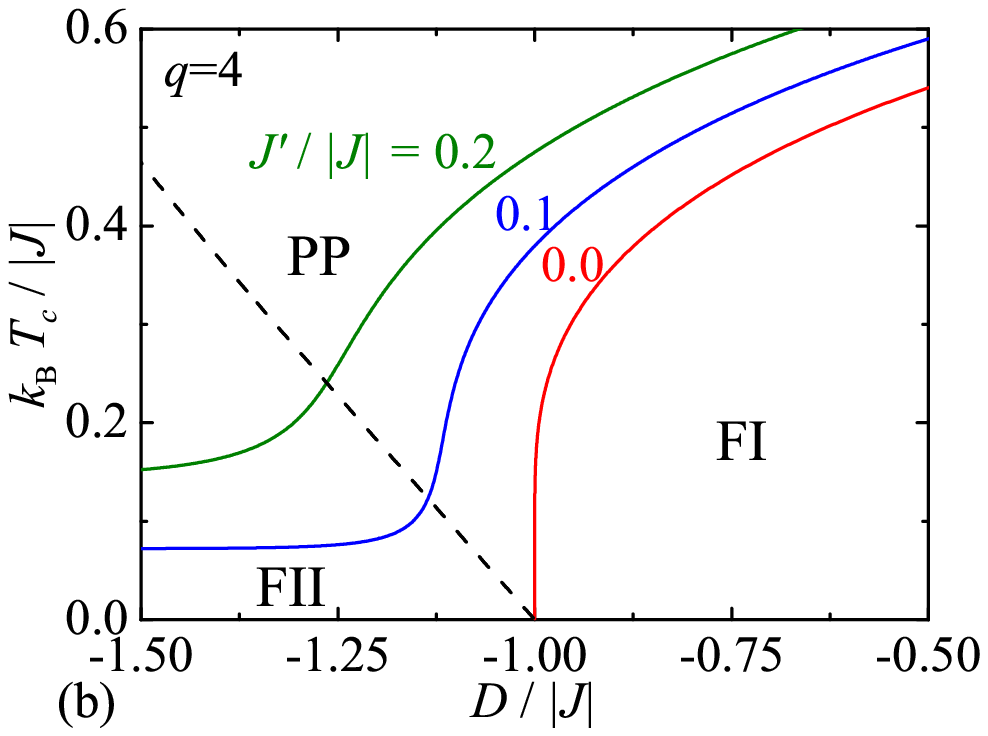}
\vspace{-1cm}
\caption{Critical temperature of the mixed spin-1/2 and spin-1 Ising model on the decorated Bethe lattice with the coordination number $q=4$ as a function of the single-ion anisotropy $D/|J|$ for several values of the ferromagnetic next-nearest-neighbour interaction $J'/|J| \geq 0$. Broken curve shows the line of compensations points and Fig. \ref{fig4}(b) is a detailed plot from the region 
near the ground-state phase boundary between FI and FII.}
\label{fig4}
\end{figure}

\begin{figure}
\includegraphics[width=0.53\textwidth]{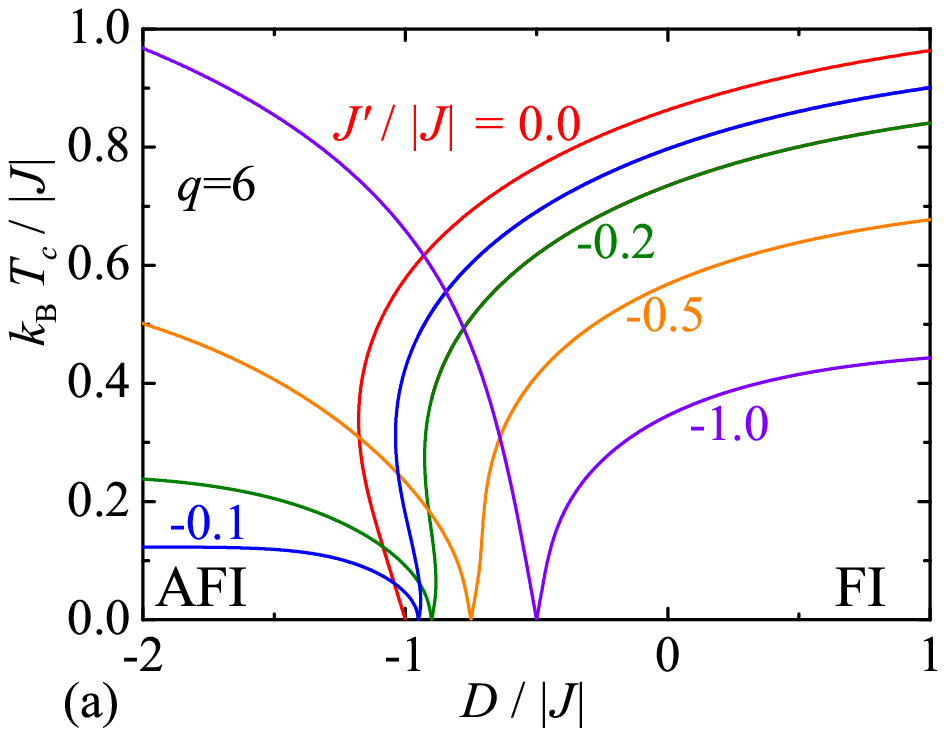}
\hspace{-0.6cm}
\includegraphics[width=0.53\textwidth]{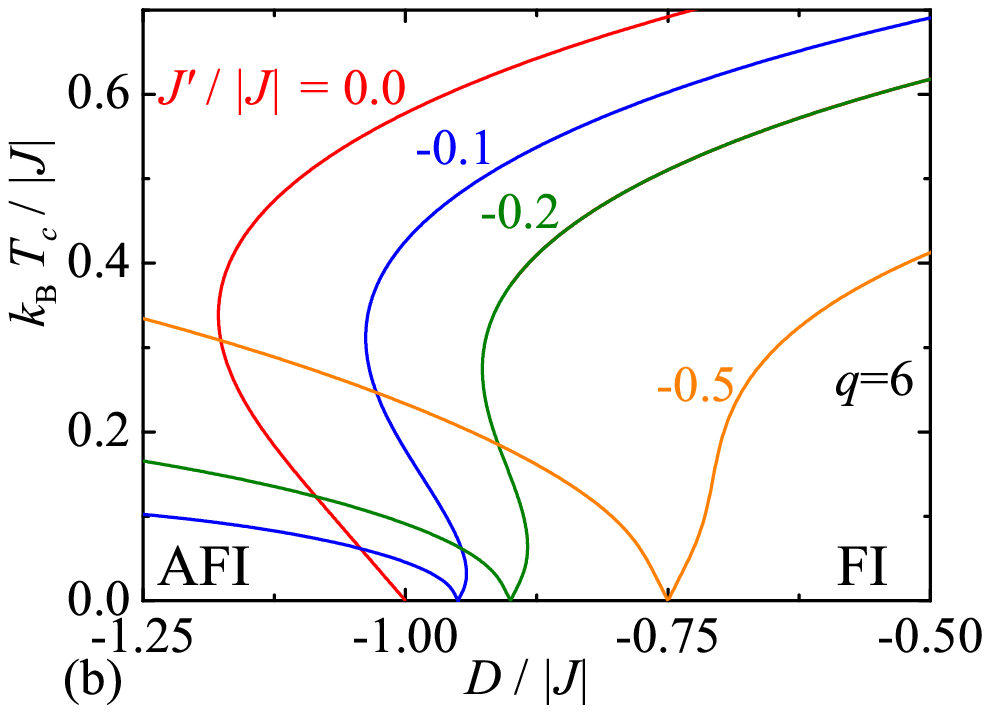}
\vspace{-1cm}
\caption{Critical temperature of the mixed spin-1/2 and spin-1 Ising model on the decorated Bethe lattice  with the coordination number $q=6$ as a function of the single-ion anisotropy $D/|J|$ for several values of the antiferromagnetic next-nearest-neighbour interaction $J'/|J| \leq 0$. Broken curve shows the line of compensations points and Fig. \ref{fig5}(b) is a detailed plot from the observed reentrant region near the ground-state phase boundary between FI and AFI.}
\label{fig5}
\end{figure}

\begin{figure}
\includegraphics[width=0.53\textwidth]{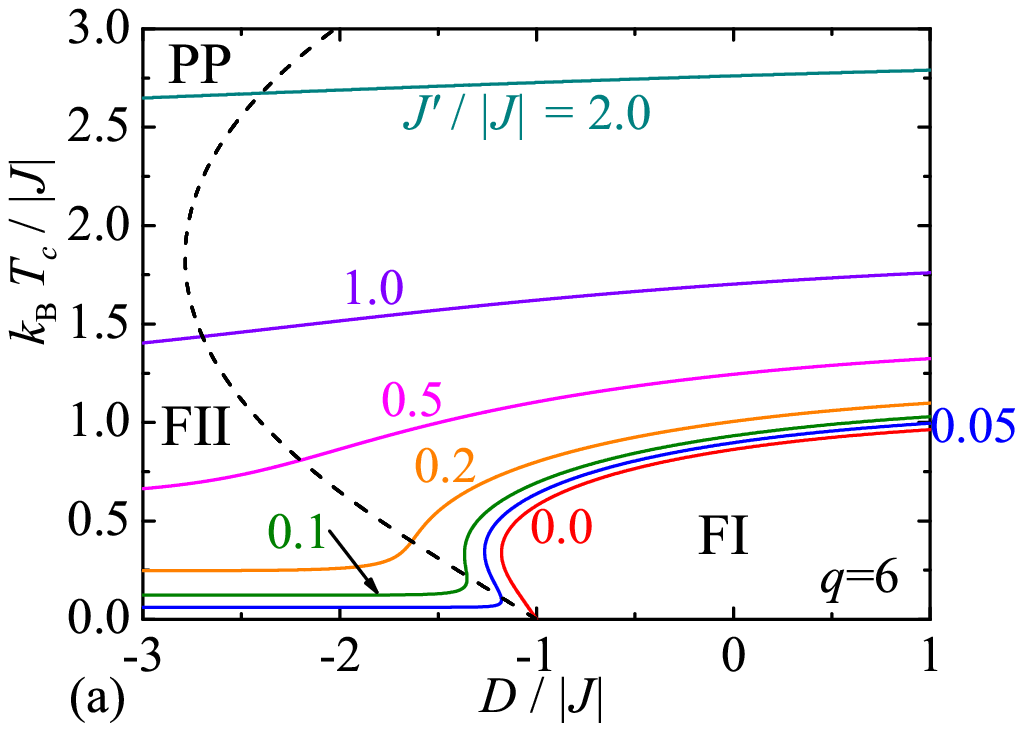}
\hspace{-0.6cm}
\includegraphics[width=0.53\textwidth]{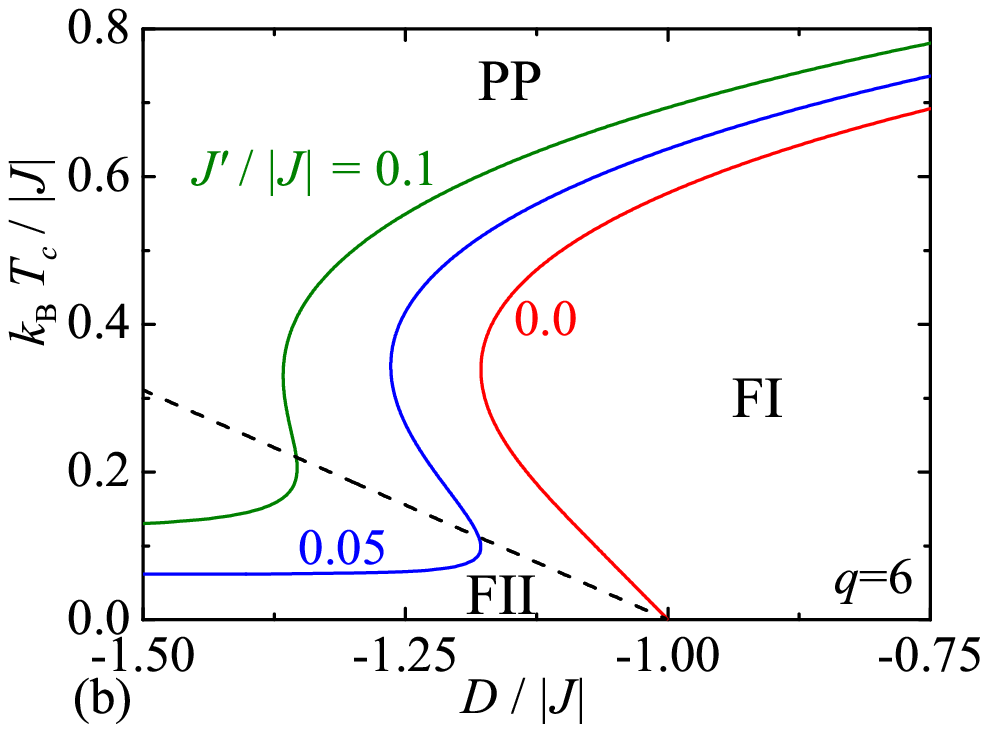}
\vspace{-1cm}
\caption{Critical temperature of the mixed spin-1/2 and spin-1 Ising model on the decorated Bethe lattice with the coordination number $q=4$ as a function of the single-ion anisotropy $D/|J|$ for several values of the ferromagnetic next-nearest-neighbour interaction $J'/|J| \geq 0$. Black broken curve shows the line of compensations points, whereas Fig. \ref{fig4}(b) depicts a detail from the region, where reentrant phase transitions can be observed.}
\label{fig6}
\end{figure}

Now, let us focus our attention on a discussion of finite-temperature phase diagrams, which are plotted in Figs. \ref{fig3}-\ref{fig6} in the form of critical temperature versus single-ion anisotropy dependences for several values of the coordination number $q$ and different strengths of the next-nearest-neighbour interaction $J'/|J|$. It is noteworthy that all displayed phase boundaries are the lines of second-order phase transitions and the ordered (disordered) phases can be located below (above) depicted critical boundaries. Fig. \ref{fig3}(a) shows the critical temperature as a function of the single-ion anisotropy in an absence of the next-nearest-neighbour interaction (i.e. $J'/|J|= 0.0$). In this particular case, the spontaneously long-range ordered FI represents the ground state for $D/|J|>-1.0$, while the disordered paramagnetic phase (PP) becomes the ground state for $D/|J|<-1.0$. As one can see, the critical line approaches the zero-temperature phase boundary between FI and PP with a negative slope for the decorated Bethe lattices with the coordination number $q<4$, with a positive slope for the decorated Bethe lattices with the coordination number $q>4$ and with an infinite gradient for the particular value of the coordination number $q=4$. These observations would suggest that reentrant phase transitions can be observed in a close vicinity of the ground-state boundary between FI and PP for the decorated Bethe lattices with a sufficiently high coordination number $q>4$, whereas the reentrance becomes the more pronounced, the higher the coordination number $q$ of the underlying Bethe lattice is. The displayed exact results may be contrasted with the approximative results reported by Kaneyoshi \cite{Kan96,Kan97}, which imply either more \cite{Kan96} or less \cite{Kan97} robust reentrant region for the decorated square lattice. On the other hand, the present results are in accordance with the previously reported exact results for mixed-spin Ising model on the decorated planar lattices, among which the decorated square \cite{Jascur98,Dakhama98}, honeycomb and kagom\'e lattice \cite{Matasovska07} do not show reentrance in contrast to the decorated triangular lattice with a pronounced reentrant region \cite{Matasovska07}. Note furthermore that the investigated model cannot exhibit the compensation phenomenon when the next-nearest-neighbour interaction is absent.

Next, we will pay our attention to the effect of the antiferromagnetic next-nearest-neighbour interaction on the critical behaviour of the mixed-spin Ising model on the decorated Bethe lattice with the coordination number $q=4$. It can be clearly seen from Fig. \ref{fig3}(b) that the critical temperature of FI monotononically decreases with decreasing the single-ion anisotropy until it reaches the zero-temperature phase boundary between FI and AFI (see the right wing of each displayed critical lines in Fig. \ref{fig3}(b)). Afterwards, the new critical line develops for a more negative values of the single-ion anisotropy, whereas the opposite trend is observed; the critical temperature monotonically increases with decreasing the single-ion anisotropy until it tends towards its maximum value in the limit $D/|J| \to - \infty$. The left wing of each critical line obviously corresponds to AFI phase with the zero spontaneous magnetization but the non-zero staggered magnetization of the spin-1/2 Ising atoms (see the subsequent discussion for more details). 
It should be noticed that the antiferromagnetic next-nearest-neighbour interaction can neither induce the reentrant phase transitions, nor the compensation phenomenon for the decorated Bethe lattices with the coordination number $q<5$. 

For completeness, the finite-temperature phase diagram of the mixed spin-1/2 and spin-1 Ising model on the decorated Bethe lattice with the coordination number $q=4$ is plotted in Fig. \ref{fig4} for different strengths of the ferromagnetic next-nearest-neighbour interaction. The critical temperature generally exhibits a smooth monotonous decline with decreasing the single-ion anisotropy and at first sight, there is no clear evidence of the phase transition between FI and FII phases in the displayed critical lines. 
It will be later demonstrated, however, that the sublattice magnetization of the spin-1 atoms approaches in the zero-temperature limit its saturation value only if $D/|J|>-1.0$, while it starts from zero in the reverse case. With this in mind, one may conclude that the critical line of FI continuously merges at some finite temperature with the critical line of FII at the specific value of the single-ion anisotropy $D/|J|=-1.0$, whereas the former FI phase is located below the right part ($D/|J|>-1.0$) and the latter FII phase below the left part ($D/|J|<-1.0$) of the overall critical line. Last but not least, it is worth mentioning that a presence of the weak (strong) ferromagnetic next-nearest-neighbour interaction may lead to an existence of one (two) compensation points in the temperature dependence of the total magnetization as it can be clearly seen from the respective behaviour of the broken curve, which displays the numerical solution of the necessary condition (\ref{eqco}) for appearance of the compensation points. It should be remembered that the necessary condition for appearance of compensation points does not depend according to Eq. (\ref{eqco}) on a relative strength of the next-nearest-neighbour interaction $J'/|J|$. From this point of view, the line of compensation temperatures starts at the ground-state boundary between FI and FII phases at $D/|J|=-1.0$ and it ends up at a respective crossing point between the necessary condition (\ref{eqco}) and the relevant critical line depending basically on a relative strength of the next-nearest-neighbour interaction (above this line the spin system is disordered and there is no spontaneous ordering). On the other hand, it also turns out that the ferromagnetic next-nearest-neighbour interaction cannot induce reentrant phase transitions for the decorated Bethe lattices with the coordination number $q<5$.

To provide a deeper insight into how the coordination number affects the overall critical behaviour, the critical temperature of the mixed spin-1/2 and spin-1 Ising model on the decorated Bethe lattice with the higher coordination number $q=6$ is plotted against the single-ion anisotropy in Figs. \ref{fig5}-\ref{fig6}. First, let us briefly comment on the most significant changes of the relevant critical lines when considering the influence of the antiferromagnetic next-nearest-neighbour interaction (Fig. \ref{fig5}). 
In the absence of the next-nearest-neighbour interaction, the investigated spin system exhibits a relatively wide reentrant region 
for the single-ion anisotropy close to but slightly below the ground-state boundary between FI and PP at $D/|J| = -1.0$.  
In this special case, the spin system is disordered in the ground state, then it orders at a lower critical temperature to display the spin order inherent to FI, which finally disappears at the upper critical temperature (PP-FI-PP reentrance). On the other hand, there are quite fundamental differences in the respective reentrant behaviour of the spin system with the antiferromagnetic next-nearest-neighbour interaction. First of all, the spin system exhibits a more striking reentrance with three consecutive critical points when starting from the spontaneously long-range ordered ground state, which either corresponds to AFI or FI depending on an interplay between the single-ion anisotropy and the next-nearest-neighbour interaction. The spontaneous ordering of AFI (or FI) disappears at the first critical temperature, then there appears PP in a relatively narrow region in between the first and second critical temperature, 
at which the spontaneous order of FI occurs (or re-appears) and finally vanishes at the third critical temperature. The observed  reentrant phase transitions can be thus characterized by two following sequences of second-order phase transformations AFI-PP-FI-PP and FI-PP-FI-PP.
It can be also clearly seen from Fig. \ref{fig5}(b) that the reentrant region moves towards a more positive values of the single-ion anisotropy (together with the ground-state boundary between FI and AFI) and it gradually vanishes upon strengthening the antiferromagnetic next-nearest-neighbour interaction.  

Finally, the last couple of phase diagrams from Fig. \ref{fig6} illustrates the effect of the ferromagnetic next-nearest-neighbour interaction on the critical behaviour of the mixed spin-1/2 and spin-1 Ising model on the decorated Bethe lattice with the coordination number $q=6$. As one can see, the character of reentrant phase transitions substantially changes by taking the ferromagnetic next-nearest-neighbour interaction into consideration even although one still detects the reentrance with three consecutive critical points. However, the observed reentrant phase transition is basically different in that both low-temperature as well as  high-temperature spontaneously long-range ordered phase is always FII. According to this, the spin system undergoes the following sequence of second-order phase transitions FII-PP-FII-PP when starting from the lowest and ending up at the highest temperature. To summarize, the only common feature of all three kinds of reentrant phase transitions FII-PP-FII-PP, FI-PP-FI-PP and AFI-PP-FI-PP is that the reentrance is quite sensitive and easily diminishes by changing a relative strength of the next-nearest-neighbour interaction.

\begin{figure}
\includegraphics[width=0.53\textwidth]{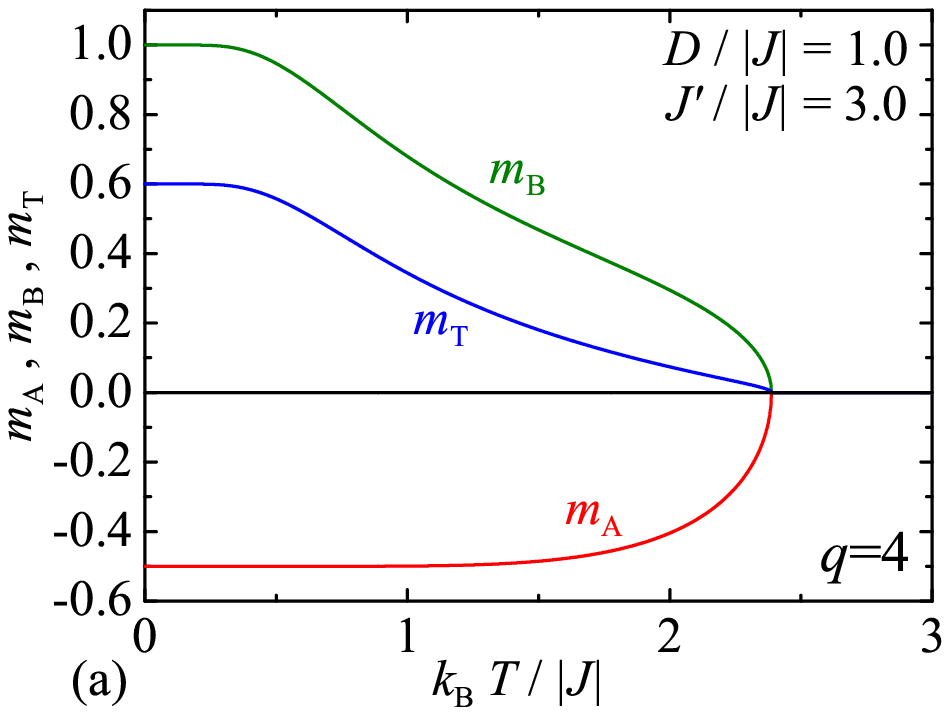}
\hspace{-0.6cm}
\includegraphics[width=0.53\textwidth]{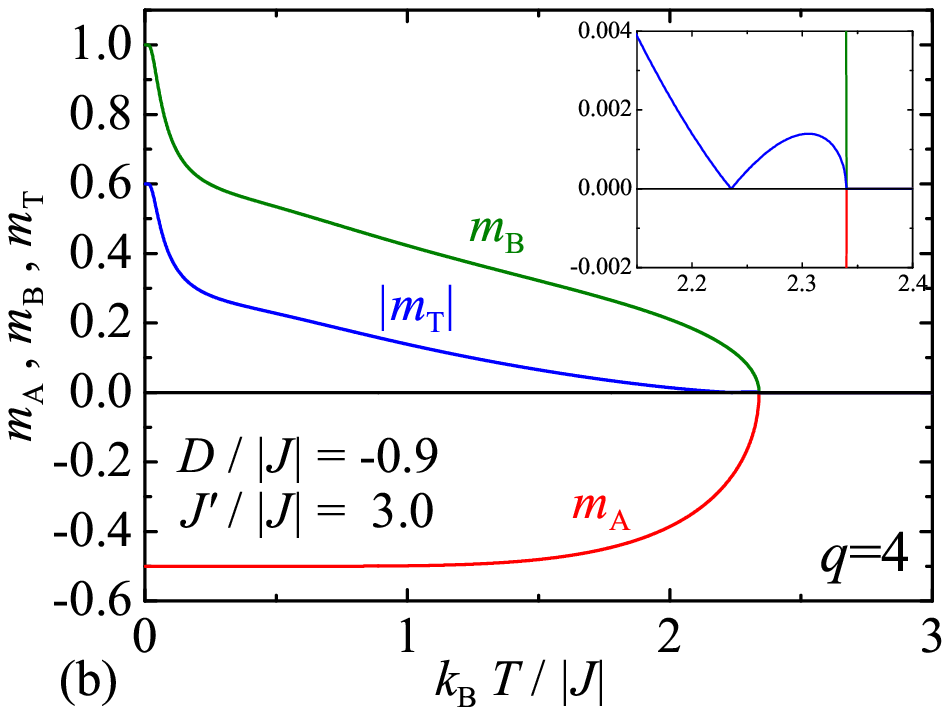}
\includegraphics[width=0.53\textwidth]{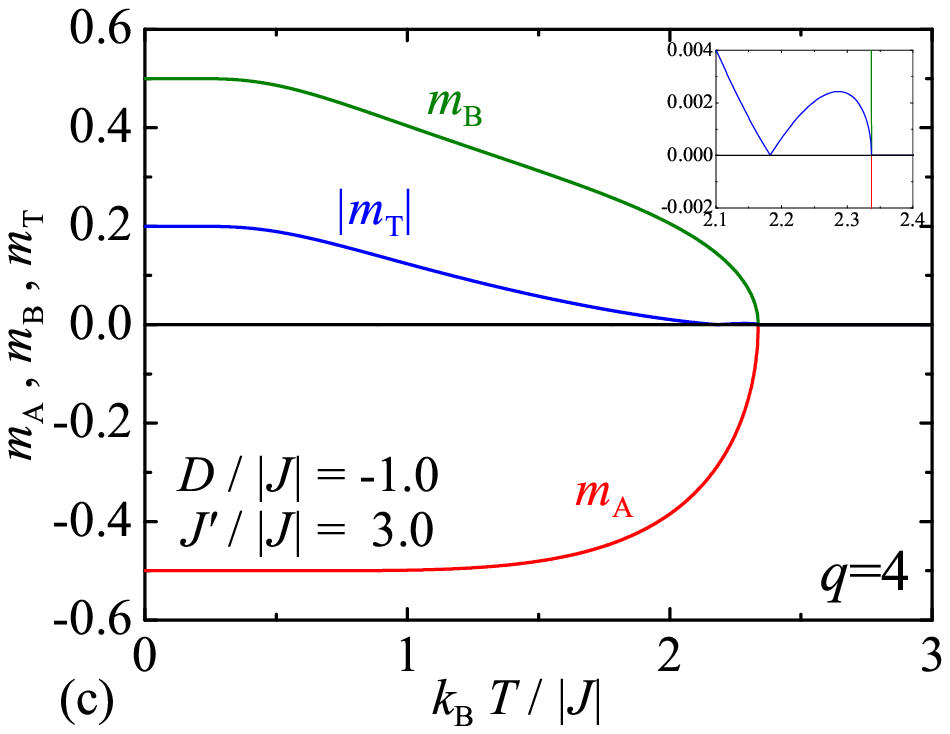}
\hspace{-0.6cm}
\includegraphics[width=0.53\textwidth]{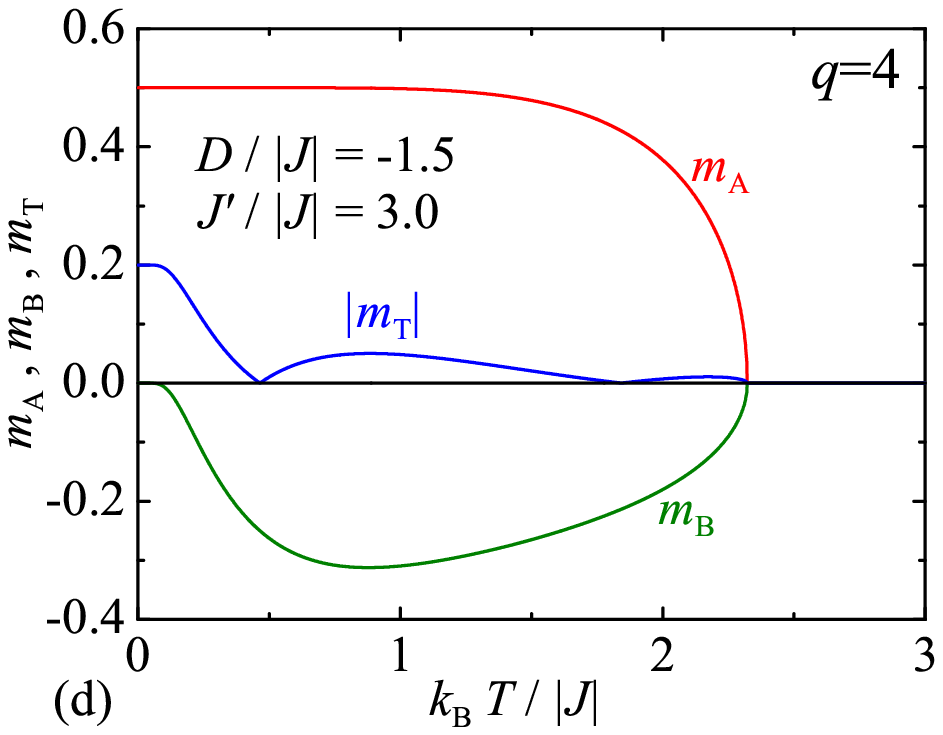}
\vspace{-1cm}
\caption{Thermal variations of the total and sublattice magnetizations for the mixed spin-1/2 and spin-1 Ising model on the decorated Bethe lattice with the coordination number $q=4$, the fixed value of the ferromagnetic next-nearest-neighbour interaction $J'/|J| = 3.0$
and four different values of the single-ion anisotropy: (a) $D/|J| = 1.0$; (b) $D/|J| = -0.9$; (c) $D/|J| = -1.0$; (b) $D/|J| = -1.5$.}
\label{fig7}
\end{figure}

\begin{figure}
\includegraphics[width=0.53\textwidth]{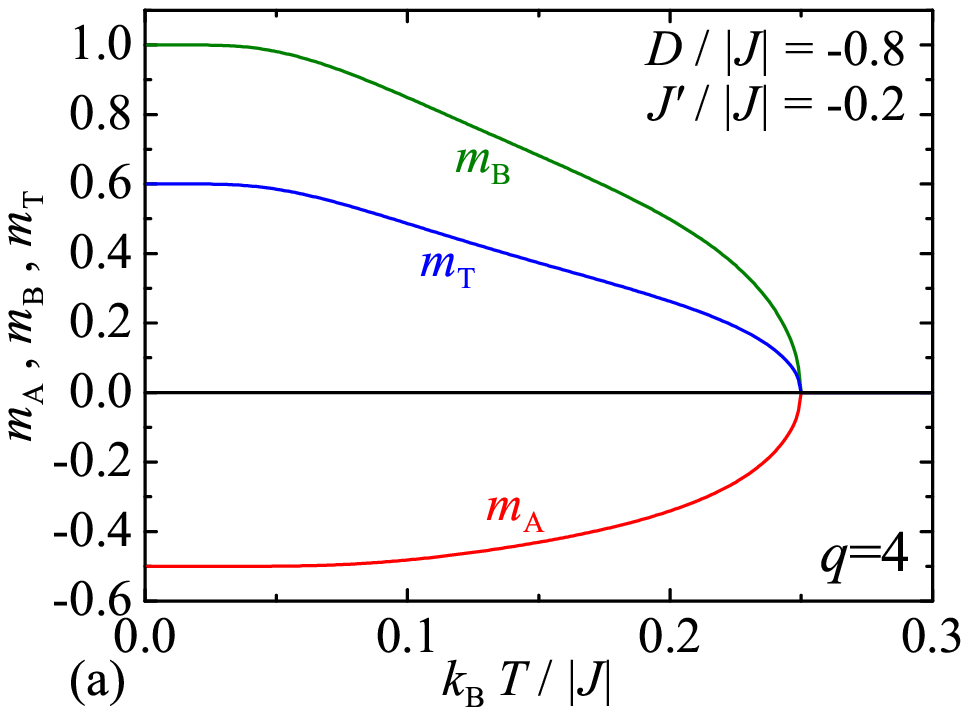}
\hspace{-0.6cm}
\includegraphics[width=0.53\textwidth]{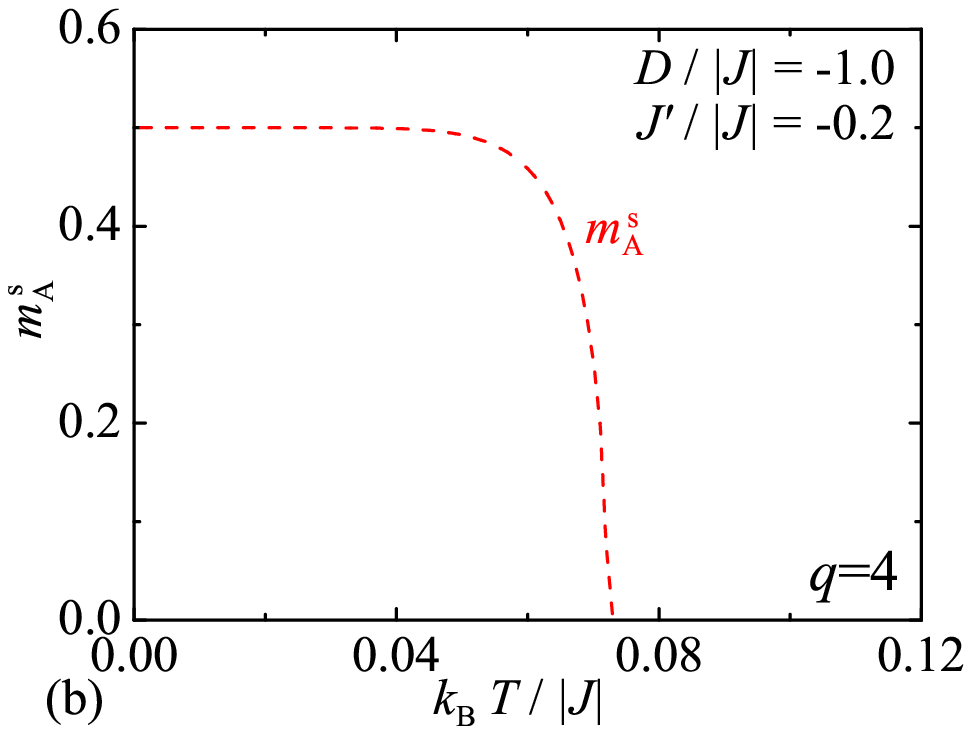}
\vspace{-1cm}
\caption{Thermal variations of the total, sublattice and staggered magnetizations for the mixed spin-1/2 and spin-1 Ising model on the decorated Bethe lattice with the coordination number $q=4$, the fixed value of the antiferromagnetic next-nearest-neighbour interaction 
$J'/|J| = -0.2$ and two different values of the single-ion anisotropy: (a) $D/|J| = -0.8$; (b) $D/|J| = -1.0$.}
\label{fig8}
\end{figure}

\begin{figure}
\includegraphics[width=0.53\textwidth]{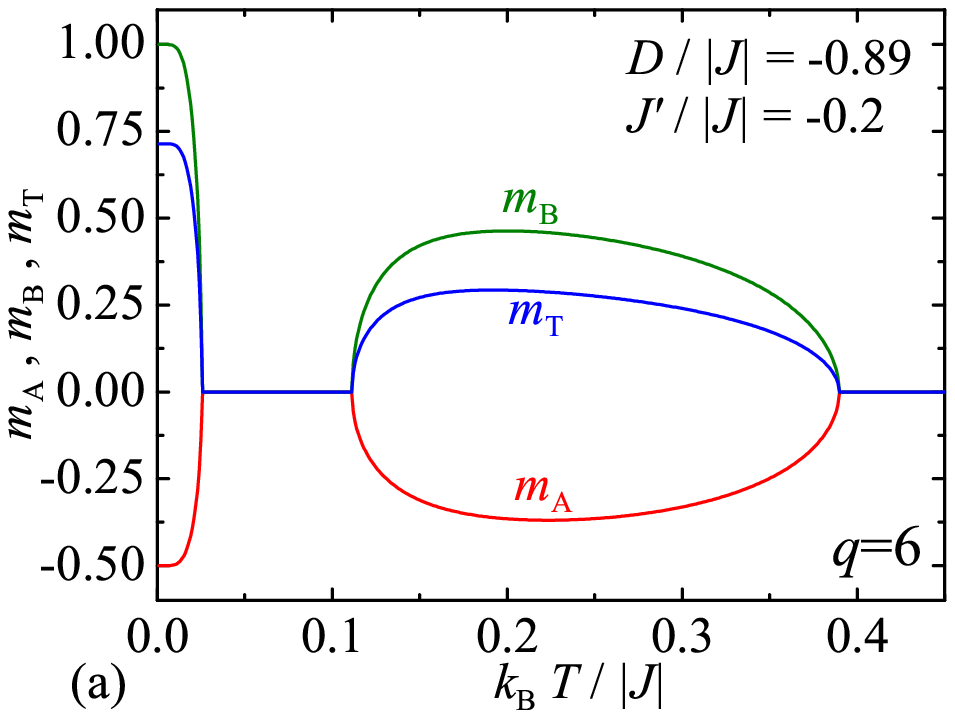}
\hspace{-0.6cm}
\includegraphics[width=0.53\textwidth]{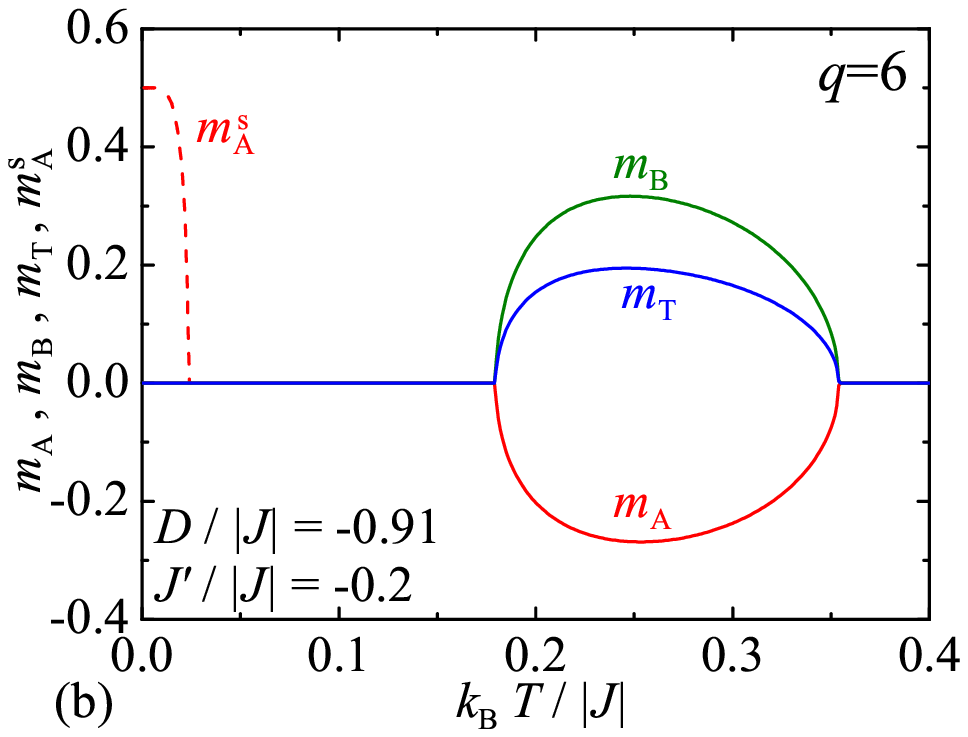}
\vspace{-1cm}
\caption{Thermal variations of the total, sublattice and staggered magnetizations for the mixed spin-1/2 and spin-1 Ising model on the decorated Bethe lattice with the coordination number $q=6$, the fixed value of the antiferromagnetic next-nearest-neighbour interaction $J'/|J| = -0.2$ and two different values of the single-ion anisotropy: (a) $D/|J| = -0.89$; (b) $D/|J| = -0.91$.}
\label{fig9}
\end{figure}

To provide an independent check of the aforedescribed critical behaviour, let us proceed to a discussion of thermal dependences of the order parameters as depicted in Figs. \ref{fig7}-\ref{fig9}. First, let us confirm the phase transition from FI towards FII by considering the constant value of the ferromagnetic next-nearest-neighbour interaction $J'/|J| = 3.0$ and changing a relative strength 
of the single-ion anisotropy. Fig. \ref{fig7}(a)--(b) illustrate some typical temperature dependences of the total and sublattice magnetizations when FI is being the ground state. The resultant magnetization exhibits for $D/|J|=1.0$ and $-0.9$ 
the standard R-type dependence and the slightly deformed N-type dependence with one compensation point, respectively, whose unusual shape 
comes from a steeper temperature-induced decrease in the sublattice magnetization $m_B$ of the decorating spin-1 atoms. Furthermore, 
Fig. \ref{fig7}(c) shows temperature variations of the total and sublattice magnetizations exactly at the phase boundary between
FI and FII. In agreement with this statement, the sublattice magnetization $m_B$ tends towards the intermediate value $0.5$ when approaching zero temperature, which indicates a coexistence of FI and FII in the ground state. Finally, thermal dependences of the total and sublattice magnetizations displayed in Fig. \ref{fig7}(d) provide an exact evidence of FII. 
As a matter of fact, the sublattice magnetization $m_B$ vanishes in the zero-temperature limit and its anomalous temperature-induced increase can be attributed to spin excitations from the non-magnetic spin state $S_i=0$ preferably towards the excited spin state 
$S_i = -1$. In the consequence of that, the resultant magnetization might display the striking W-type dependence with two successive compensation points not included in the standard N\'eel theory of ferrimagnetism (for the characterization of temperature dependences of the total magnetization see Refs. \cite{jozef06,cesur09}). The former compensation point appears owing to a higher number of the decorating spins (the overall contribution of the sublattice magnetization $m_B$ to the total magnetization consequently exceeds the one of the sublattice magnetization $m_A$ provided that $m_B$ becomes sufficiently high), while the latter compensation point emerges as a result of more subtle character of the decorating spins with respect to thermal fluctuations (the sublattice magnetization $m_B$ tends steeper to zero than the sublattice magnetization $m_A$ in a vicinity of the critical point).

Next, let us turn to thermal variations of the total and sublattice magnetizations of the mixed spin-1/2 and spin-1 Ising model on the decorated Bethe lattice with the coordination number $q=4$ and the antiferromagnetic next-nearest-neighbour interaction $J'/|J| = -0.2$. It has been demonstrated that the antiferromagnetic next-nearest-neighbour interaction competes with the nearest-neighbour interaction and hence, the weaker easy-plane single-ion anisotropy is thus generally needed in order to destroy the spontaneous order of FI. However, one still detects temperature dependences of the total and sublattice magnetizations quite typical for FI whenever the single-ion anisotropy is selected above its critical value $D_b/|J| = -1-J'/(2|J|)$ (see Fig. \ref{fig8}(a)). Contrary to this, both spontaneous sublattice magnetizations equal zero for any single-ion anisotropy below this boundary value $D<D_b$ and the system surprisingly exhibits the unusual antiferromagnetic long-range order AFI with the non-zero staggered magnetization of the spin-1/2 atoms as the relevant order parameter (see Fig. \ref{fig8}(b)). Hence, it actually turns out that the spin-1/2 Ising atoms from nearest-neighbour shells of the decorated Bethe lattice align antiparallel with respect to each other due to the antiferromagnetic next-nearest-neighbour interaction and owing to this fact, the decorating spin-1 atoms become frustrated and all tend towards their non-magnetic spin state $S_i=0$. 

For completeness, let us also confirm an existence of the reentrant phase transitions by analyzing thermal dependences of the spontaneous magnetization of the mixed spin-1/2 and spin-1 Ising model on the decorated Bethe lattice with the higher coordination number $q=6$.
For this purpose, we depict in Fig. \ref{fig9} temperature variations of the total and sublattice magnetizations for one specific value 
of the antiferromagnetic next-nearest-neighbour interaction $J'/|J| = -0.2$ and two different values of the single-ion anisotropy.
In the former case with $D/|J| = -0.89$, one observes three successive reentrant phase transitions just in between two phases FI and PP in the following order FI-PP-FI-PP. On the other hand, another type of reentrance can be detected in the latter case with $D/|J| = -0.89$, which reveals three consecutive phase transitions between AFI, FI and PP in the following order AFI-PP-FI-PP. It should be stressed that the latter reentrance would be misinterpreted as the reentrant phase transition with only two consecutive critical points if the staggered magnetization would not be taken into account as the possible order parameter along with the uniform magnetization.  

\section{Conclusion}

In this work, we have exactly calculated magnetic properties of the mixed spin-1/2 and spin-$S$ Ising model on the decorated Bethe lattice with a quite general coordination number $q$ when combining the decoration-iteration transformation with the exact recursion relations. The model Hamiltonian includes the nearest-neighbour interaction between the spin-1/2 and spin-$S$ atoms, the next-nearest-neighbour interaction between the spin-1/2 atoms and the uniaxial single-ion anisotropy. By making use of the decoration-iteration transformation, we have established a precise mapping equivalence between the mixed spin-1/2 and spin-$S$ Ising model on the decorated Bethe lattice and its corresponding spin-1/2 Ising model on the simple (undecorated) Bethe lattice subsequently treated with the aid of exact recursion relations. Furthermore, the rigorous mapping theorems and exact spin identities have been employed in order to get the precise results for both spontaneous sublattice magnetizations as well as staggered magnetization. 

Our main attention was focused on magnetic properties of the mixed spin-1/2 and spin-1 Ising ferrimagnet, which exhibits remarkably diverse critical behaviour and compensation phenomena. The four different phases have been found in the ground-state phase diagram, two of them have the uniform magnetization and another two staggered magnetization as the order parameter. In addition, it has been  demonstrated that the investigated mixed-spin system may exhibit either reentrant phase transitions with two consecutive critical points (for zero next-nearest-neighbour interaction) or three different types of reentrant phase transitions with three successive critical points (for non-zero next-nearest-neighbour interaction). It is worthy to recall that the reentrance may appear just for a sufficiently high coordination number $q>4$ of the underlying Bethe lattice and this observation seems to be of a quite general validity as it has been already proved in the Ising-Heisenberg models on the diamond-like decorated Bethe lattices \cite{jozef10,cesur11}. Finally, it has been also illustrated that the next-nearest-neighbour interaction may cause a presence of one or two compensation points in the temperature dependence of the total magnetization.  

Let us conclude our study by mentioning that the present formalism can be rather straightforwardly extended for the mixed spin-1/2 and spin-$S$ Ising model on the decorated Bethe lattices in a presence of the external magnetic field and this exactly soluble model might display a quite intriguing magnetization process including several non-trivial magnetization plateaus. In this direction will continue our further efforts.

\end{document}